# Approximate capacity of the two-way relay channel: A deterministic approach

Amir Salman Avestimehr[1], Aydin Sezgin[2] and David N.C. Tse[1]
[1]Wireless Foundations, UC Berkeley, California, USA.
[2]Stanford University, Information Systems Laboratory, California, USA
email:avestime@eecs.berkeley.edu, sezgin@stanford.edu, dtse@eecs.berkeley.edu

*Abstract*—We study the capacity of the full-duplex bidirectional (or two-way) relay channel with two nodes and one relay. The channels in the forward direction are assumed to be different (in general) than the channels in the backward direction, i.e. channel reciprocity is not assumed. We use the recently proposed deterministic approach to capture the essence of the problem and to determine a good transmission and relay strategy for the Gaussian channel. Depending on the ratio of the individual channel gains, we propose to use either a simple amplify-and-forward or a particular superposition coding strategy at the relay. We analyze the achievable rate region and show that the scheme achieves to within 3 bits the cut-set bound for all values of channel gains.

## I. INTRODUCTION

Bidirectional or two-way communication between two nodes was first studied by Shannon himself in [1]. Nowadays the two-way communication where an additional node acting as a relay is supporting the exchange of information between the two nodes is attracting increasing attention. Some achievable rate regions for the two-way relay channel using different strategies at the relay, such as decode-and-forward, compress-and-forward and amplify-and-forward, have been analyzed in [2]. The capacity region of the so called broadcast two-way half-duplex relay channel, i.e. assuming that the communication takes places in two hops and the relay is decoding the received messages completely, was recently characterized in [3]. Network coding type techniques have been proposed by [4], [5], [6] (and others) in order to improve the transmission rate. While inferior to traditional routing at low signal-to-noise-ratios (SNR), it was shown in [7] that network coding achieves twice the rate of routing at high SNR. Similarly, in [8] the half-duplex two-way relay channel where the channel gains are all equal to one is investigated. It was shown that a combination of a decode-and-forward strategy using lattice codes and a joint decoding strategy is asymptotically optimal. Indeed, by using lattice codes it was shown in [9] that for some cases rates within less than one bit to the capacity can be achieved.

So far, the main focus is however so far on the one-way relay channel, which was introduced by [10] and further investigated in [11]. In general, cooperative communication schemes are particulary important when reliable communication can not be guaranteed by using a conventional point-to-point connection. Cooperation between two source nodes for communication to a common receiver was proposed in [12]. There, a non-cooperative phase is followed by a cooperative one and it is shown that this strategy outperforms non-cooperative strategies. Cooperation by using distributed space-time coding techniques in networks has been analyzed in [13], [14], [15], [16]. Recent information-theoretic studies on relay channels can be found in e.g. [17] and references therein. Relaying can be expected to be adopted in current and future wireless systems, as it has been introduced in the 802.16j (WiMAX) standard.

In this paper, we study the capacity of the full-duplex two-way relay channel, which, to the best of our knowledge, is not known in general. Motivated by the deterministic approach in [18] for Gaussian networks, here we make progress towards the goal of "approximating" the capacity region of the two-way relay channel. The advantage of the deterministic approach is that one can focus on the interaction between the signals arriving from different nodes rather than the background noise of the system. Thus, our work represents an alternative approach, however for the full-duplex case, to e.g. the approaches in [2], [4], [8]. Furthermore, here we analyze the general case, where the channel gains are all different (in general) and channel reciprocity is not assumed. Although our focus is on the case where a direct link between the two nodes is not present, we discuss also the impact of a direct channel later on. Similar to the general relay network studied in [18], [19] and the interference channel studied in [20], [21], we show that our scheme can achieve to within 3 bits of the capacity for all channel parameter values.

## II. SYSTEM MODEL

The system model of the two-way full-duplex relay channel is shown in Fig. 1. Communication takes place simultaneously from the relay to the nodes and vice versa. As can be observed from Fig. 1, channel reciprocity is not assumed here. Thus, in general $h_1$, which is the channel parameter describing the link from node $A$ to the relay, is different from $h_3$, the channel describing the link from the relay to node $A$ (and similarly for $h_2$ and $h_4$). The received signal at the relay is given by (cf.Fig. 1(a))

$$y_R = h_1 x_A + h_2 x_B + z_R, \quad (1)$$

where $x_A$ and $x_B$ are the signals transmitted from node $A$ and node $B$, respectively. The variable $z_R$ describes the additive

The work of A.Sezgin is supported by the Deutsche Forschungsgemeinschaft (DFG).

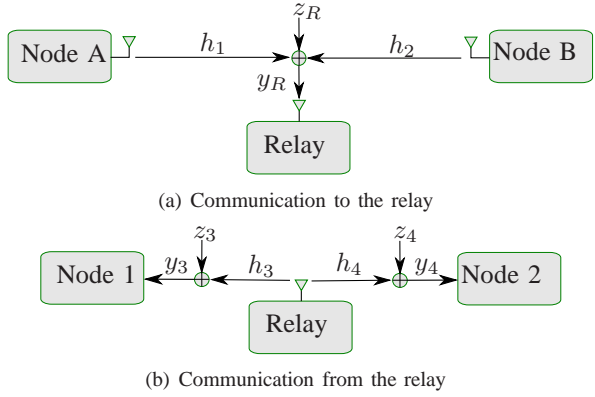

Fig. 1. Bidirectional relaying

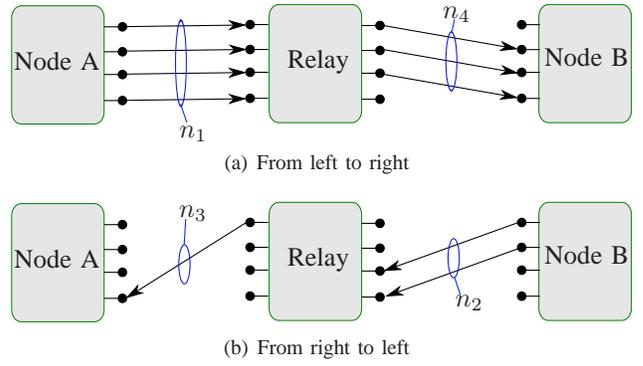

Fig. 2. Deterministic model for bidirectional relaying

Gaussian noise at the relay. Without loss of generality, we assume that $\mathbb{E}\left[|x_A|^2\right] = \mathbb{E}\left[|x_B|^2\right] = \mathbb{E}\left[|z_R|^2\right] = 1$. The received signals at the nodes are given by (cf. Fig. 1(b))

$$y_A = h_3 x_R + z_2 \quad (2)$$
$$y_B = h_4 x_R + z_3.$$

The variables $z_2$ and $z_3$ are the unit variance additive Gaussian noises at node $A$ and node $B$, respectively.

## III. DETERMINISTIC TWO-WAY RELAY

The deterministic channel model was introduced in [18]. Here is a formal definition of this model.

**Definition 1: (Definition of the deterministic model)** Consider a wireless network as a set of nodes $V$, where $|V| = N$. Communication from node $i$ to node $j$ has a non-negative integer gain[1] $n_{(i,j)}$ associated with it. This number models the channel gain in a corresponding Gaussian setting. At each time $t$, node $i$ transmits a vector $\mathbf{x}_i[t] \in \mathbb{F}_2^q$ and receives a vector $\mathbf{y}_i[t] \in \mathbb{F}_2^q$ where $q = \max_{i,j}(n_{(i,j)})$. The received signal at each node is a deterministic function of the transmitted signals at the other nodes, with the following input-output relation: if the nodes in the network transmit $\mathbf{x}_1[t], \mathbf{x}_2[t], \ldots \mathbf{x}_N[t]$ then the received signal at node $j$, $1 \leq j \leq N$ is:

$$\mathbf{y}_j[t] = \sum_{k=1}^{N} \mathbf{S}^{q-n_{k,j}} \mathbf{x}_k[t] \quad (3)$$

for all $1 \leq k \leq N$, where $\mathbf{S}$ is the $q \times q$ shift matrix given by

$$\mathbf{S} = \begin{bmatrix} 0 & 0 & 0 & \cdots & 0 \\ 1 & 0 & 0 & \cdots & 0 \\ 0 & 1 & 0 & \cdots & 0 \\ \vdots & \ddots & \ddots & \ddots & \vdots \\ 0 & \cdots & 0 & 1 & 0 \end{bmatrix}.$$

and the summation and multiplication is in $\mathbb{F}_2$.

We start our analysis by considering the deterministic model of the two-way relay channel as shown in Fig. 2. The following theorem is our main result for the deterministic two-way relay network.

[1]Some channels may have zero gain.

**Theorem 1:** The capacity region of the bi-directional linear finite field deterministic relay network is:

$$R_{AB} \leq \min(n_1, n_4) \quad (4)$$
$$R_{BA} \leq \min(n_2, n_3). \quad (5)$$

Furthermore, the cut-set bound is achievable with a simple shift-and-forward strategy at the relay.

In the rest of the section, we give a sketch of the proof. We use an algebraic approach to solve the problem of finding the optimal strategy. In the deterministic model assume that node $A$ and $B$ sends $\mathbf{x}_A$ and $\mathbf{x}_B \in \mathbb{F}_2^q$, respectively, where $q = \max(n_1, n_2, n_3, n_4)$. The received signal at the relay is then given by

$$\mathbf{y}_R = \mathbf{S}^{q-n_1}\mathbf{x}_A + \mathbf{S}^{q-n_2}\mathbf{x}_B.$$

Now consider a linear coding strategy at the relay. Thus, it is going to send

$$\mathbf{x}_R = \mathbf{G}\mathbf{y}_R = \mathbf{G}(\mathbf{S}^{q-n_1}\mathbf{x}_A + \mathbf{S}^{q-n_2}\mathbf{x}_B),$$

where $\mathbf{G}$ is an arbitrary $q \times q$ generating matrix that is a design choice.

The received signal at node $A$ is thus given by

$$\mathbf{y}_A = \mathbf{S}^{q-n_3}\mathbf{x}_R = \mathbf{S}^{q-n_3}\mathbf{G}(\mathbf{S}^{q-n_1}\mathbf{x}_A + \mathbf{S}^{q-n_2}\mathbf{x}_B)$$

while node $B$ receives

$$\mathbf{y}_B = \mathbf{S}^{q-n_4}\mathbf{x}_R = \mathbf{S}^{q-n_4}\mathbf{G}(\mathbf{S}^{q-n_1}\mathbf{x}_A + \mathbf{S}^{q-n_2}\mathbf{x}_B).$$

Since node $A$ and node $B$ respectively know their own signals $\mathbf{x}_A$ and $\mathbf{x}_B$, they can cancel it from their received signal. Hence effectively they receive

$$\mathbf{y}'_A = \mathbf{S}^{q-n_3}\mathbf{G}\mathbf{S}^{q-n_2}\mathbf{x}_B$$
$$\mathbf{y}'_B = \mathbf{S}^{q-n_4}\mathbf{G}\mathbf{S}^{q-n_1}\mathbf{x}_A. \quad (6)$$

The question is, whether we can find a matrix $\mathbf{G}$, such that the rates $R_{AB} = \min(n_1, n_4)$ and $R_{BA} = \min(n_2, n_3)$ in (4), (5) are achievable. By obtaining such a matrix, we would also gain insights how the processing at the relay should be done in an optimal way.

Now we state the following lemma, whose proof is given in Appendix A.

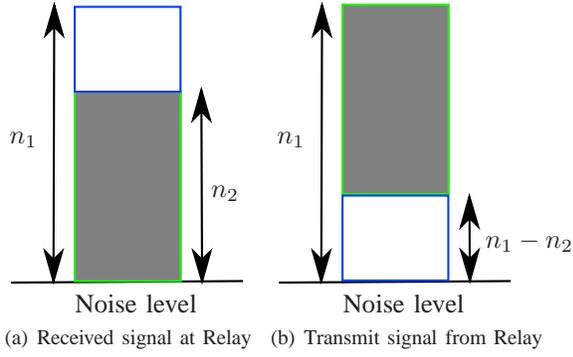

(a) Received signal at Relay  (b) Transmit signal from Relay

Fig. 3. Signal levels at relay: Receive phase and transmit phase

**Lemma 1:** It is possible to convert the network in Fig. 2 into one of the following two cases without changing the cut-set bound.
1) $n_1 = n_4$ and $n_2 \leq n_3$
2) $n_2 = n_3$ and $n_1 \leq n_4$

Therefore, by Lemma 1 and symmetry we only need to study this case:
$$n_1 = n_4 \text{ and } n_2 \leq n_3.$$

It turns out, that we are indeed able to construct a matrix **G**, such that the cut-set bound is achievable. The generating matrices **G** for the individual cases (the derivation is given in Appendix B) are given as follows.

1) $q = n_1$
$$\mathbf{G} = \begin{bmatrix} \mathbf{0}_{n_2 \times (q-n_2)} & \mathbf{I}_{n_2} \\ \mathbf{I}_{q-n_2} & \mathbf{0}_{(q-n_2) \times n_2} \end{bmatrix} \quad (7)$$

2) $q = n_3$
   (a) $n_2 \leq n_1$
$$\mathbf{G} = \begin{bmatrix} \mathbf{0}_{n_1 \times (q-n_1)} & \mathbf{I}_{n_1} \\ \mathbf{0}_{q-n_1} & \mathbf{0}_{(q-n_1) \times n_1} \end{bmatrix} \quad (8)$$
   (b) $n_2 > n_1$
$$\mathbf{G} = \begin{bmatrix} \mathbf{0}_{n_1 \times (q-n_1)} & \mathbf{I}_{n_1} \\ \mathbf{I}_{q-n_1} & \mathbf{0}_{(q-n_1) \times n_1} \end{bmatrix} \quad (9)$$

In the following we give interpretations of the different generating matrices **G** in (7), (8), and (9) for the three cases.

*A. Interpretation of the case $n_1 = q$*

We start with the generating matrix **G** in (7). The interpretation of this operation for the deterministic case is the following. The relay receives $n_1 = q$ signal levels as shown in Fig. 3(a). The last $n_2$ contain information from both node $A$ and node $B$ (gray area in Fig. 3(a)) and the other (top) signal levels are only information from $A$ (white area in Fig. 3(a)). The relay is now creating a codeword, which has the last $n_2$ received signal levels at highest level (gray area in Fig. 3(b)) and the remaining bits of $A$ at lower signal levels (white area in Fig. 3(a)).

The interpretation of the scheme for the Gaussian channel is the following. At first the relay decodes a part of the message, namely $\mathbf{x}_A^{(1)}$, received from node $A$ that has arrived above the signal level of node $B$ and subtracts it from the overall received signal. The remaining part (lowest $n_2$ levels) of the overall received signal at the relay is just the summation of signals from both the node $A$ and node $B$. The argumentation here is that the relay can not decode this summation and thus it quantize it. The interesting part is now that the relay creates the transmit signal by using a superposition code [22]. The cloud center of this superposition code is the quantized signal, while the bin index is the information $\mathbf{x}_A^{(1)}$ it has decoded from node $A$.

*B. Interpretation of the case $n_3 = q$*

We start with the case $n_2 \leq n_1$. Here, the relay receives $n_1$ signal levels. The relay then simply shifts the received signal up and forwards it. The corresponding scheme for the Gaussian channel is thus amplify-forward. As an alternative approach, we could also use a similar superposition strategy as for $n_1 = q$. However, as we will show later on, the simple amplify-and-forward strategy is enough in order to achieve to within 3 bits the capacity for all channel parameter values.

The case with $n_2 > n_1$ is analogous to the case with $n_1 = q$. Here the relay receives $n_2$ signal levels. The last $n_1$ bits contain information for both node $A$ and $B$ and the rest is just the information for node $A$. The interpretation of the scheme for the Gaussian channel is very similar to the scheme for $n_1 = q$ and thus omitted.

## IV. GAUSSIAN TWO-WAY RELAY CHANNEL

In this section, we use the insights obtained from studying the deterministic two-way relay channel to find near-optimal relaying strategies in the Gaussian case as defined in section II. It follows our main result for the Gaussian two-way relay channel and the rest of this section is devoted to proving it.

**Theorem 2:** Consider a Gaussian two-way relay channel as defined in section II with unit average noise and transmit power at each node. The capacity of this system satisfies
$$\bar{C}_{AB} - 3 \leq C_{AB} \leq \bar{C}_{AB}$$
and
$$\bar{C}_{BA} - 3 \leq C_{BA} \leq \bar{C}_{BA},$$
where $\bar{C}_{AB} = \log(1 + \min(|h_1|^2, |h_4|^2))$ and $\bar{C}_{BA} = \log(1 + \min(|h_2|^2, |h_3|^2))$ is the cut-set upper bound on the capacity of the transmission from $A$ to $B$ and $B$ to $A$, respectively [23].

Since Lemma 1 holds also for the Gaussian case, we again need to study only the case that $|h_1|^2 = |h_4|^2$ and $|h_2|^2 \leq |h_3|^2$. Now we discuss the achievability strategy:

*A. Achievability strategy*

In general, the transmit signals from node $A$, node $B$ and the relay are given by
$$\mathbf{x}_A = \sqrt{\alpha_A} \mathbf{x}_A^{(1)} + \sqrt{1 - \alpha_A} \mathbf{x}_A^{(2)}$$
$$\mathbf{x}_B = \sqrt{\alpha_B} \mathbf{x}_B^{(1)} + \sqrt{1 - \alpha_B} \mathbf{x}_B^{(2)}$$
$$\mathbf{x}_R = \sqrt{\alpha_R} \mathbf{x}_R^{(1)} + \sqrt{1 - \alpha_R} \mathbf{x}_R^{(2)}. \quad (10)$$

where $\mathbf{x}_A^{(1)}$, $\mathbf{x}_A^{(2)}$, $\mathbf{x}_B^{(1)}$, $\mathbf{x}_B^{(2)}$, $\mathbf{x}_R^{(1)}$, and $\mathbf{x}_R^{(2)}$ are codewords chosen from a random Gaussian codebook of size $2^{nR_{AB}^{(1)}}$, $2^{nR_{AB}^{(2)}}$, $2^{nR_{BA}^{(1)}}$, $2^{nR_{BA}^{(2)}}$, $2^{nR_R^{(1)}}$, and $2^{nR_R^{(2)}}$, respectively. At node $A$ (and similarly for node $B$) we have two messages $\mathbf{m}_A^{(1)}$ and $\mathbf{m}_A^{(2)}$ of size $2^{nR_{AB}^{(1)}}$ and $2^{nR_{AB}^{(2)}}$ that are mapped to $x_A^{(1)}$ and $x_A^{(2)}$, respectively. The relay signaling strategy depends on the channel gains and will be specificized later for each case. The choice of $\alpha_A$, $\alpha_B$, and $\alpha_R$ in (10) depend on the magnitude of the channel gains $|h_1|$, $|h_2|$, $|h_3|$, and $|h_4|$.

### B. $|h_1|^2 \geq |h_3|^2$

Following the insights gained from the deterministic model, for $|h_1|^2 \geq |h_3|^2$ we set $\alpha_B = 0$ and $R_{BA}^{(1)} = 0$. The transmit signal at node $B$ then reduces to

$$\mathbf{x}_B = \mathbf{x}_B^{(2)}.$$

Thus, the receive signal at the relay is given by

$$\mathbf{y}_R = \left(\sqrt{\alpha_A}\mathbf{x}_A^{(1)} + \sqrt{1-\alpha_A}\mathbf{x}_A^{(2)}\right)h_1 + h_2\mathbf{x}_B + \mathbf{z}_R. \quad (11)$$

$\alpha_A$ is chosen such that the received signal of $\mathbf{x}_A^{(2)}$ and $\mathbf{x}_B$ are at the same scale. Thus, the following expression has to hold

$$\sqrt{1-\alpha_A}h_1 = h_2, \quad (12)$$

which gives

$$1 - \alpha_A = \left(\frac{h_2}{h_1}\right)^2.$$

Form $\mathbf{y}_R$, the relay first decodes $\mathbf{x}_A^{(1)}$ (i.e. $\mathbf{m}_A^{(1)}$) by treating the remaining received signals $\mathbf{x}_A^{(2)}$ and $\mathbf{x}_B$ as noise. This can be done with low error probability as long as

$$R_{AB}^{(1)} \leq \log\left(1 + \frac{\alpha_A|h_1|^2}{1+(1-\alpha_A)|h_1|^2+|h_2|^2}\right)$$
$$= \log\left(1 + \frac{|h_1|^2-|h_2|^2}{1+2|h_2|^2}\right). \quad (13)$$

Then the relay maps the decoded $\mathbf{x}_A^{(1)}$ to another codeword $\mathbf{x}_R^{(1)}$ of size $2^{nR_R^{(1)}}$ with $R_R^{(1)} = R_{AB}^{(1)}$. If the above expression is fulfilled, the relay can decode the signal $\mathbf{x}_A^{(1)}$ and cancel it from the received signal in (11). Thus, we have

$$\tilde{\mathbf{y}}_R = \sqrt{1-\alpha_A}\mathbf{x}_A^{(2)}h_1 + h_2\mathbf{x}_B + \mathbf{z}_R.$$

As suggested in the deterministic model, $\tilde{\mathbf{y}}_R$ is not decoded. Rather, a quantization is performed. The relay uses an optimal vector quantizer of size $2^{nR_R^{(2)}}$ and maps the quantization index to a codeword $\mathbf{x}_R^{(2)}$. Then the relay transmits (10), where

$$\alpha_R = \frac{\alpha_A}{2|h_2|^2+1}.$$

Having received the signal from the relay, nodes $A$ and $B$ first attempt to decode $\mathbf{x}_R^{(2)}$. Since node $A$ knows $\mathbf{x}_R^{(1)}$ it can cancel it from the received signal, however node $B$ is treating $\mathbf{x}_R^{(1)}$ as noise. The decoding of $\mathbf{x}_R^{(2)}$ can be done with low error probability as long as

$$R_R^{(2)} \leq \min\left(\log\left(1 + \frac{|h_1|^2(1-\alpha_R)}{|h_1|^2\alpha_R+1}\right), \log\left(1+|h_3|^2(1-\alpha_R)\right)\right) \quad (14)$$
$$= \min\left(\log\left(\frac{|h_1|^2+1}{|h_1|^2\alpha_R+1}\right), \log\left(1+|h_3|^2(1-\alpha_R)\right)\right).$$

The second expression within the min-operation is obtained due to node $A$. As aforementioned, assuming that node $A$ knows the strategy of relay and the codebook it has used, it can reconstruct $\mathbf{x}_R^{(1)}$ perfectly, since it contains only its own message. Using interference cancelation results in a interference free channel. The first expression within the min-operation is obtained due to node $B$ which observes part of the signal from the relay, i.e. $\mathbf{x}_R^{(1)}$, as additional noise. Then node $B$ cancels $\mathbf{x}_R^{(2)}$ from its received signal and attempts to decode $\mathbf{x}_R^{(1)}$. This can be done with low error probability if

$$R_R^{(1)} \leq \log\left(1 + \alpha_R|h_1|^2\right).$$

Now that nodes $A$ and $B$ have decoded $\mathbf{x}_R^{(1)}$, they can create

$$\tilde{\mathbf{y}}_R^Q = \beta\tilde{\mathbf{y}}_R + \mathbf{z}_Q = \beta\left(\sqrt{1-\alpha_A}\mathbf{x}_A^{(2)}h_1 + h_2\mathbf{x}_B + \mathbf{z}_R\right) + \mathbf{z}_Q$$
$$\stackrel{(12)}{=} \beta\left(h_2\left(\mathbf{x}_A^{(2)}+\mathbf{x}_B\right)+\mathbf{z}_R\right)+\mathbf{z}_Q$$

where

$$\beta = (1-D/\sigma_{\tilde{\mathbf{y}}_R}^2)$$

and $\mathbf{z}_Q$ is due to the quantization noise with variance

$$\sigma_Q^2 = D(1-D/\sigma_{\tilde{\mathbf{y}}_R}^2).$$

Thus, the distortion $D$ in our case has to fulfill [23]

$$D = 2^{-R_R^{(2)}}\sigma_{\tilde{y}_R}^2 = \min\left(\frac{\alpha_R|h_1|^2+1}{|h_1|^2+1}, \frac{1}{1+|h_3|^2(1-\alpha_R)}\right)$$
$$\times \left(2|h_2|^2+1\right).$$

Assuming that the nodes are able to cancel their own message from $\tilde{\mathbf{y}}_R^Q$, they can decode each others codeword with low error probability if

$$R_{BA} \leq \min\left(\log\left(1 + \frac{|h_2|^2\left(1-\frac{\alpha_R|h_1|^2+1}{|h_1|^2+1}\right)}{1+\frac{\alpha_R|h_1|^2+1}{|h_1|^2+1}2|h_2|^2}\right), \right. \quad (15)$$
$$\left. \log\left(1 + \frac{|h_2|^2\left(1-\frac{1}{1+|h_3|^2(1-\alpha_R)}\right)}{1+\frac{2|h_2|^2}{1+|h_3|^2(1-\alpha_R)}}\right)\right)$$

and

$$R_{AB}^{(2)} \leq \min\left(\log\left(1 + \frac{|h_2|^2\left(1-\frac{\alpha_R|h_1|^2+1}{|h_1|^2+1}\right)}{1+\frac{\alpha_R|h_1|^2+1}{|h_1|^2+1}2|h_2|^2}\right), \right. \quad (16)$$
$$\left. \log\left(1 + \frac{|h_2|^2\left(1-\frac{1}{1+|h_3|^2(1-\alpha_R)}\right)}{1+\frac{2|h_2|^2}{1+|h_3|^2(1-\alpha_R)}}\right)\right).$$

Therefore, the rate in (15) and

$$R_{AB} \stackrel{(13),(16)}{\leq} \log\left(1 + \frac{|h_1|^2 - |h_2|^2}{1 + 2|h_2|^2}\right)$$
$$+ \min\left(\log\left(1 + \frac{|h_2|^2\left(1 - \frac{\alpha_R|h_1|^2+1}{|h_1|^2+1}\right)}{1 + \frac{\alpha_R|h_1|^2+1}{|h_1|^2+1}2|h_2|^2}\right),\right.$$
$$\left.\log\left(1 + \frac{|h_2|^2\left(1 - \frac{1}{1+|h_3|^2(1-\alpha_R)}\right)}{1 + \frac{2|h_2|^2}{1+|h_3|^2(1-\alpha_R)}}\right)\right)$$

are achievable. With some algebra, we can show that

$$\min\left(\log\left(1 + \frac{|h_2|^2\left(1 - \frac{\alpha_R|h_1|^2+1}{|h_1|^2+1}\right)}{1 + \frac{\alpha_R|h_1|^2+1}{|h_1|^2+1}2|h_2|^2}\right),\right.$$
$$\left.\log\left(1 + \frac{|h_2|^2\left(1 - \frac{1}{1+|h_3|^2(1-\alpha_R)}\right)}{1 + \frac{2|h_2|^2}{1+|h_3|^2(1-\alpha_R)}}\right)\right)$$
$$\geq \log\left(1 + |h_2|^2\right) - \log(3)$$

and

$$\log\left(1 + \frac{|h_1|^2 - |h_2|^2}{1 + 2|h_2|^2}\right)$$
$$+ \min\left(\log\left(1 + \frac{|h_2|^2\left(1 - \frac{\alpha_R|h_1|^2+1}{|h_1|^2+1}\right)}{1 + \frac{\alpha_R|h_1|^2+1}{|h_1|^2+1}2|h_2|^2}\right),\right.$$
$$\left.\log\left(1 + \frac{|h_2|^2\left(1 - \frac{1}{1+|h_3|^2(1-\alpha_R)}\right)}{1 + \frac{2|h_2|^2}{1+|h_3|^2(1-\alpha_R)}}\right)\right)$$
$$\geq \log\left(1 + |h_1|^2\right) - \max(2,3).$$

Thus, we are at most 3 bits away from the cut-set bound.

### C. Case $|h_1|^2 < |h_3|^2$

*1) Amplify-and-forward:* $|h_2|^2 < |h_1|^2$: With $\alpha_A = \alpha_B = 0$, the transmit signals from node $A$ and node $B$ reduce to $\mathbf{x}_A = \mathbf{x}_A^{(2)}$ and $\mathbf{x}_B = \mathbf{x}_B^{(2)}$ chosen from a random Gaussian codebook of size $2^{nR_{AB}}$ and $2^{nR_{BA}}$, respectively. Thus, the received signal at the relay is given by

$$\mathbf{y}_R = h_1\mathbf{x}_A + h_2\mathbf{x}_B + \mathbf{z}_R.$$

Using a amplify and forward strategy, the transmit signal at the relay is thus given by

$$\mathbf{x}_R = \frac{1}{\sqrt{|h_1|^2 + |h_2|^2 + 1}}\mathbf{y}_R.$$

Using (2), the received signals at the nodes are given by

$$\mathbf{y}_A = \frac{h_3}{\sqrt{|h_1|^2 + |h_2|^2 + 1}}(h_1\mathbf{x}_A + h_2\mathbf{x}_B + \mathbf{z}_R) + \mathbf{z}_A$$
$$\mathbf{y}_B = \frac{h_4}{\sqrt{|h_1|^2 + |h_2|^2 + 1}}(h_1\mathbf{x}_A + h_2\mathbf{x}_B + \mathbf{z}_R) + \mathbf{z}_B.$$

First, the nodes cancel their own messages from the received signal. Then, the nodes can decode each other signals with low error probability as long as

$$R_{AB} \leq \log\left(1 + \frac{|h_1|^2|h_4|^2}{|h_4|^2 + |h_1|^2 + |h_2|^2 + 1}\right)$$

$$R_{BA} \leq \log\left(1 + \frac{|h_2|^2|h_3|^2}{|h_3|^2 + |h_1|^2 + |h_2|^2 + 1}\right).$$

With some algebra, we can show that

$$\log\left(1 + \frac{|h_1|^2|h_4|^2}{|h_4|^2 + |h_1|^2 + |h_2|^2 + 1}\right) \geq \log\left(1 + |h_1|^2\right) - \log(3)$$

$$\log\left(1 + \frac{|h_2|^2|h_3|^2}{|h_3|^2 + |h_1|^2 + |h_2|^2 + 1}\right) \geq \log\left(1 + |h_2|^2\right) - \log(3).$$

Thus, we are at most within $\log(3)$ bits away from the cut-set bound, which is strictly better than what we aimed for.

*2)* $|h_2|^2 > |h_1|^2$: The following derivations are very similar to the case $|h_1|^2 \geq |h_3|^2$ with slight differences. First of all, $\alpha_A = 0$ and $\alpha_B$ and $\alpha_R$ are now given by

$$\alpha_B = 1 - \frac{|h_1|^2}{|h_2|^2} \text{ and } \alpha_R = \frac{\alpha_B|h_2|^2}{|h_3|^2\left(2|h_1|^2 + 1\right)}.$$

While we had a min-operator in the case $|h_1|^2 > |h_3|^2$ (cf. (14), here it can be shown that $|h_1|^2 > |h_3|^2/(\alpha_R|h_3|^2+1)$ is never fulfilled in this case. Thus, we have to consider only $|h_1|^2 \leq |h_3|^2/(|h_3|^2\alpha_R+1)$ and the min-operator is obsolete. Therefore the nodes can decode each other signals with low error probability as long as

$$R_{AB} \leq \log\left(1 + \frac{|h_1|^2\left(1 - \frac{1}{1+|h_1|^2(1-\alpha_R)}\right)}{1 + \frac{2|h_1|^2}{1+|h_1|^2(1-\alpha_R)}}\right)$$

and

$$R_{BA} \leq \log\left(\frac{1 + |h_1|^2 + |h_2|^2}{1 + 2|h_1|^2}\right) + R_{AB}.$$

With some algebra, we can show that

$$\log\left(1 + \frac{|h_1|^2\left(1 - \frac{1}{1+|h_1|^2(1-\alpha_R)}\right)}{1 + \frac{2|h_1|^2}{1+|h_1|^2(1-\alpha_R)}}\right) \geq \log\left(1 + |h_1|^2\right) - \log(3)$$

and

$$\log\left(\frac{1 + |h_1|^2 + |h_2|^2}{1 + 2|h_1|^2}\right) + R_{AB} \geq \log\left(1 + |h_2|^2\right) - 3$$

Thus, we are at most 3 bits away from the cut-set bound.

## V. IMPACT OF A DIRECT LINK BETWEEN NODES

If a direct link between the nodes $A$ and $B$ is present then the system equations change to

$$y_A = h_3 x_R + h_5 x_B + z_2 \qquad (17)$$
$$y_B = h_4 x_R + h_6 x_A + z_3.$$

Since channel reciprocity is not assumed, in general $h_5 \neq h_6$. The cut-set bound for the deterministic case changes to

$$R_{AB} \leq \min(\max(n_1, n_6), \max(n_4, n_6))$$
$$= n_6 + \min((n_1 - n_6)^+, (n_4 - n_6)^+) \qquad (18)$$
$$R_{BA} \leq \min(\max(n_2, n_5), \max(n_3, n_5))$$
$$= n_5 + \min((n_2 - n_5)^+, (n_3 - n_5)^+). \qquad (19)$$

From the cut-set bound above, we observe that as long as $n_5$ and $n_6$ are larger than $\max(n_1, n_2, n_3, n_4)$, we can ignore the relay. If that is not the case, then the relay ignores the top $q-\max(\min((n_2-n_5)^+,(n_3-n_5)^+),\min((n_1-n_6)^+,(n_4-n_6)^+))$, signal levels, with $q=\max(n_1,\ldots,n_6)$, at the relay. Then, the first $\min((n_2-n_5)^+,(n_3-n_5)^+,(n_1-n_6)^+,(n_4-n_6)^+)$ are routed from the nodes over the relay at an interfering signal level. The intermediate $|\min((n_2-n_5)^+,(n_3-n_5)^+)-\min((n_1-n_6)^+,(n_4-n_6)^+)|$ are routed over the relay on the non-interfering signal levels.

For the Gaussian two-way relay channel we have the following cut-set bound

$$\bar{C}_{AB} = \max_{|\rho_A|\leq 1} \min\Big( \log(1+(1-\rho_A^2)(|h_6|^2+|h_1|^2)),$$
$$\log(1+|h_6|^2+|h_4|^2+2\rho_A|h_6||h_4|)\Big)$$

$$\bar{C}_{BA} = \max_{|\rho_B|\leq 1} \min\Big( \log(1+(1-\rho_B^2)(|h_5|^2+|h_2|^2)),$$
$$\log(1+|h_5|^2+|h_3|^2+2\rho_B|h_5||h_3|)\Big).$$

The simultaneous transmission from the relay and the nodes causes interference at the respective receiving node. If $|h_6| < \min(|h_1|,|h_4|)$ or $|h_5| < \min(|h_2|,|h_3|)$, using a simple block-Markov encoding scheme in combination with backward decoding in order to overcome the interference created by the two incoming signals at each node results in the same rates for the proposed scheme as before. A better exploitation of the direct link would certainly result in higher rates. Similarly, if $|h_6| > \min(|h_1|,|h_4|)$ and $|h_5| > \min(|h_2|,|h_3|)$, i.e. the direct links are stronger than the relay links, the relay can not increase the capacity by more than 2 bits. Thus, we can ignore it and still we are within a constant gap to the cut-set bound. The cut-set bound in the interesting case in which the direct links are weaker than the relay paths can be upper bounded by

$$\bar{C}_{AB} \leq \min\big(\log(1+|h_1|)+1, \log(1+|h_4|^2)+2\big)$$
$$= \log(1+|h_1|)+1$$
$$\bar{C}_{BA} \leq \min\big(\log(1+|h_2|)+1, \log(1+|h_3|^2)+2\big)$$
$$= \log(1+|h_2|)+1.$$

Since the cut-set bound increases to at most one more bit, we are at most 4 bits away from the cut-set bound.

## VI. ILLUSTRATION

In Fig. 4(a) and 4(b), the gap between the rates $R_{AB}$ and $R_{BA}$ and the corresponding cut-set upper bound is plotted for different channel gains, respectively. The $x$-coordinate is representing the ratio of the channel gain from the relay to node $A$ (i.e. $h_3$) to the reverse direction, i.e. from node $A$ to the relay (i.e. $h_1$), in dB scale. On the $y$-coordinate we have the ratio of the channel and from the node $B$ to the relay (i.e. $h_2$) to the reverse direction, i.e. from the relay to node $B$ (i.e. $h_4 = h_1$), in dB scale. The ordinate shows the gap in bits. From the simulations, we observe that the gap is in general less than 3 bits, which verifies our theoretical results. We also observe that for a certain region, the gap is less than 1 bit. This region is especially large for $R_{BA}$. In the plot, we normalized the channel gain $h_1$ to 20 dB higher than the noise variance.

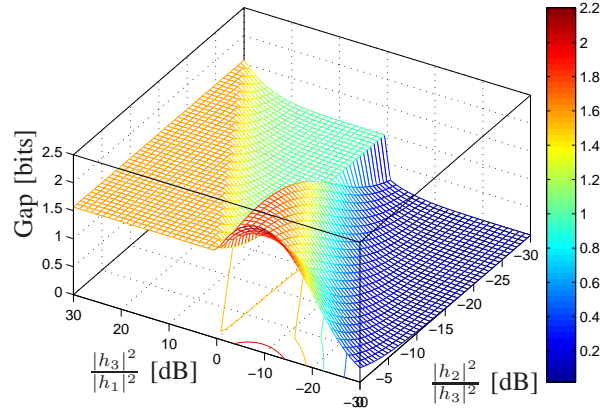

(a) Gap for $R_{AB}$

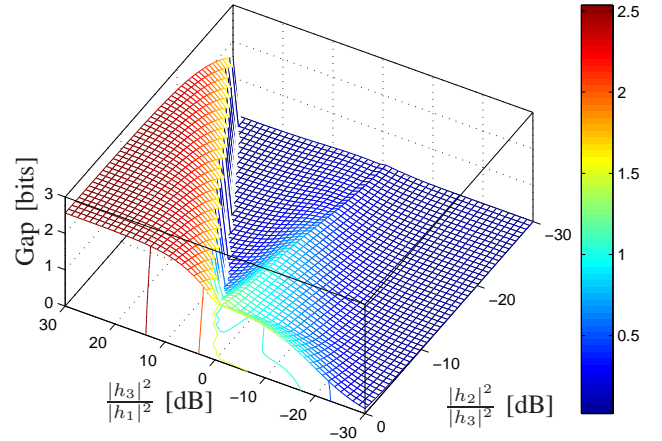

(b) Gap for $R_{BA}$

Fig. 4. Gap to the cut-set upper bound

Interestingly, it turns out that the gap is further reduced by shrinking the channel gain $h_1$ (not shown here).

## VII. CONCLUSION

In this paper, we have studied the capacity of the full-duplex bidirectional (or two-way) relay channel with two nodes and one relay. We used the recently proposed deterministic approach to capture the essence of the problem and to determine a good transmission and relay strategy for the Gaussian channel. Depending on the ratio of the individual channel gains, we used either a simple amplify-and-forward or a particular superposition coding strategy at the relay. We analyzed the achievable rate region and showed that the scheme achieves to within 3 bits the cut-set bound for all values of channel gains.

## APPENDIX A
## PROOF OF LEMMA1

*Proof:* The basic idea is that by reducing the transmission power at the nodes appropriately, the cut-set bound is not changed. Note the following three observations:

- If $n_1 > n_4$ then node $A$ can reduce its power such that $n_1' = n_4$ and also since during this process $\min(n_1, n_4)$ is unchanged, the cut-set upper bound also does not change
- Similarly if $n_2 > n_3$ then node $B$ can reduce its power such that $n_2' = n_3$ and also since during this process $\min(n_2, n_3)$ is unchanged, the cut-set upper bound also does not change
- If $n_1 < n_4$ and $n_2 < n_3$ then the relay can reduce its power by $\min(n_4-n_1, n_3-n_2)$ and then either ($n_3' = n_2$ and $n_1 \leq n_4'$) or ($n_1 = n_4'$ and $n_2 \leq n_3'$)

Therefore in any case we can transform the network to one of the cases described above, and the cut-set upper bound is not changed. ∎

## APPENDIX B
### DERIVATION OF THE GENERATING MATRICES

From (6) note that for any $\mathbf{G}$ it holds that

$$\text{rank}(\mathbf{S}^{q-n_3}\mathbf{G}\mathbf{S}^{q-n_2}) \leq \min(n_2, n_3) \quad (20)$$
$$\text{rank}(\mathbf{S}^{q-n_4}\mathbf{G}\mathbf{S}^{q-n_1}) \leq \min(n_1, n_4),$$

since from [24] we have that

$$\text{rank}(\mathbf{S}^{q-n_3}\mathbf{G}\mathbf{S}^{q-n_2}) \leq \min(\text{rank}(\mathbf{S}^{q-n_3}), \text{rank}(\mathbf{G}),$$
$$\text{rank}\mathbf{S}^{q-n_2}))$$
$$\text{rank}(\mathbf{S}^{q-n_4}\mathbf{G}\mathbf{S}^{q-n_1}) \leq \min(\text{rank}(\mathbf{S}^{q-n_4}), \text{rank}(\mathbf{G}),$$
$$\text{rank}(\mathbf{S}^{q-n_1})),$$

which is consistent with the cut-set upper bound in (4), (5).

What remains to be solved is to find a matrix $\mathbf{G}$ such that both necessary conditions in (20) are satisfied with equality, i.e.

$$\text{rank}(\mathbf{S}^{q-n_3}\mathbf{G}\mathbf{S}^{q-n_2}) = \min(n_2, n_3) \quad (21)$$
$$\text{rank}(\mathbf{S}^{q-n_4}\mathbf{G}\mathbf{S}^{q-n_1}) = \min(n_1, n_4). \quad (22)$$

Now we have two possibilities. Either we have $q = n_1$ or $q = n_3$. If $q = n_1$ then conditions on $\mathbf{G}$ in (21) and (22) become:

$$\text{rank}(\mathbf{S}^{q-n_3}\mathbf{G}\mathbf{S}^{q-n_2}) = \min(n_2, n_3) = n_2$$
$$\text{rank}(\mathbf{G}) = q$$

One $\mathbf{G}$ that satisfies both equalities is the following:

$$\mathbf{G} = \begin{bmatrix} \mathbf{0}_{n_2 \times (q-n_2)} & \mathbf{I}_{n_2} \\ \mathbf{I}_{q-n_2} & \mathbf{0}_{(q-n_2) \times n_2} \end{bmatrix}.$$

If $q = n_3$ then conditions on $\mathbf{G}$ in (21) and (22) become:

$$\text{rank}(\mathbf{G}\mathbf{S}^{q-n_2}) = \min(n_3, n_3) = n_2$$
$$\text{rank}(\mathbf{S}^{q-n_1}\mathbf{G}\mathbf{S}^{q-n_1}) = n_1$$

Here, we have to consider two cases: If $n_2 \leq n_1$ holds, then one $\mathbf{G}$ that satisfies both inequalities is the following:

$$\mathbf{G} = \begin{bmatrix} \mathbf{0}_{n_1 \times (q-n_1)} & \mathbf{I}_{n_1} \\ \mathbf{0}_{q-n_1} & \mathbf{0}_{(q-n_1) \times n_1} \end{bmatrix}.$$

If instead $n_2 > n_1$ is given, then one $\mathbf{G}$ that satisfies both inequalities is the following:

$$\mathbf{G} = \begin{bmatrix} \mathbf{0}_{n_1 \times (q-n_1)} & \mathbf{I}_{n_1} \\ \mathbf{I}_{q-n_1} & \mathbf{0}_{(q-n_1) \times n_1} \end{bmatrix}.$$